\documentclass[onecollarge,natbib]{svjour2}
\bibpunct{[}{]}{;}{n}{}{,} 
\smartqed  
\usepackage{graphicx}
\journalname{Archive of Applied Mechanics}
\usepackage{color}
\def\Journal#1#2#3#4{{#1} {\bf#2}, #3 (#4)}

\def\NPA{{\rm Nucl. Phys.} A}
\def\NPB{{\rm Nucl. Phys.} B}
\def\PLB{{\rm Phys. Lett.}  B}

\def\PRD{{\rm Phys. Rev.} D}

\def\la{\langle}
\def\ra{\rangle}

\def\lam{\lambda}

\def\be{\begin{equation}}
\def\ee{\end{equation}}
\def\bea{\begin{eqnarray}}
\def\eea{\end{eqnarray}}
\begin{document}

\title{Twist-3 Distribution Amplitudes of Pion in the Light-Front Quark Model
\thanks{This work was supported by the Korean Research Foundation Grant funded by the Korean Government
(No. NRF-2014R1A1A2057457).}
}


\author{Ho-Meoyng   Choi      \and
        Chueng-Ryong Ji 
}


\institute{H.-M. Choi \at
              Department of Physics, Teachers College, Kyungpook National University, Daegu, Korea 41566 \\
               \email{homyoung@knu.ac.kr}           
           \and
           C.-R. Ji \at
              Department of Physics, North Carolina State University, Ralegh, NC 27695-8202\\
              \email{crji@ncsu.edu}
}

\date{Received: date / Accepted: date}

\maketitle

\begin{abstract}
We analyzed two twist-3 distribution amplitudes of pion, i.e. pseudoscalar $\phi^P_{3;\pi}(x)$
and pseudotensor $\phi^\sigma_{3;\pi}(x)$, within the LFQM. Our LFQM descriptions 
both for twist-3 $\phi^P_{3;\pi}$  and  $\phi^\sigma_{3;\pi}$ obtained from the
Gaussian radial wave function
not only satisfy the fundamental constraint required from the isospin symmetry, but also
reproduce exactly the asymptotic forms anticipated from QCD's conformal limit.
\end{abstract}

\section{Introduction}
\label{intro}
Hadronic light-cone distribution amplitudes~(DAs)
have been known to play an essential role in
the QCD description of hard exclusive processes via the factorization theorem~\cite{BL80}.
Taking the transverse separation to zero, the factorization takes the form
of a convolution of a perturbatively calculable hard-scattering amplitude
and the process-independent nonperturbative DAs.
These nonperturbative DAs motivated many theoretical studies to calculate meson DAs using nonperturbative methods
such as the QCD sum rule~\cite{BF,BBL,MPS10}, the chiral-quark
model from the instanton vacuum~\cite{Goeke,NK06}, the Dyson-Schwinger
equations (DSE) approach~\cite{DSE13,DSE2015}, and the light-front quark model~(LFQM)~\cite{CJ_DA,CJ_V14,CJ2015,FB2015}.
Among them, the LFQM appears to be one of the most efficient and effective tools in studying hadron physics as
it takes advantage of the distinguished features of the  light-front dynamics (LFD)~\cite{BPP}.
Specifically, the rational energy-momentum dispersion relation  of LFD, namely
$p^-=({\bf p}^2_\perp + m^2) / p^+$, yields the sign correlation between the  LF energy $p^-(=p^0-p^3)$ and the LF
longitudinal momentum $p^+(=p^0 + p^3)$ and leads to the suppression of vacuum fluctuations in LFD.
This simplification is a remarkable advantage in LFD and facilitates the partonic interpretation of the amplitudes. Based on the advantage of LFD, the LFQM
has been quite successful in describing various static and non-static
properties of hadrons such as meson mass spectra~\cite{CJ_99}, the decay constants~\cite{Choi07},
electromagnetic and weak transition form factors~\cite{CJ_Bc} and
generalized parton distribution~(GPDs)~\cite{CJ_GPD01}.

In Ref.~\cite{CJ_DA}, we have analyzed twist-2
DAs of pseudoscalar~($\phi^A_{2;M}(x)$) and vector~($\phi^{||}_{2;V}(x)$) mesons
using the LFQM~\cite{CJ_99}.
In more recent works~\cite{CJ_V14,CJ2015}, we have extended our LFQM to analyze
twist-3 pseudoscalar ($\phi^P_{3;M}(x)$) DAs of
pseudoscalar mesons~\cite{CJ2015} and chirality-even
twist-3 ($\phi^{\perp}_{3;V}(x)$) DAs of vector mesons~\cite{CJ_V14} and
discussed the link between the chiral symmetry of QCD and the numerical results of the LFQM.
In particular,
we have discussed a wave function dependence of the LF
zero-mode~\cite{Zero1,Zero2,Zero3}
contributions to $\phi^P_{3;M}(x)$ and $\phi^{\perp}_{3;V}(x)$
{\color{black} not only for the exactly solvable
manifestly covariant Bethe-Salpeter (BS) model but also} for the more phenomenologically
accessible realistic LFQM~\cite{CJ_DA,CJ_99} using the standard LF
(SLF) approach.
We also  linked the covariant BS model to the standard LFQM {\color{black} providing}
the correspondence relation~\cite{CJ_V14,CJ2015} between the two
models. {\color{black} Effectively, we prescribed a consistent substitution for} the LF vertex function in the covariant
BS model with the more phenomenologically accessible
Gaussian wave function provided by the LFQM analysis of
meson mass~\cite{CJ_99}. The remarkable finding is that the zero-mode
contribution as well as the instantaneous contribution
revealed in the covariant BS model become absent in the
LFQM. Without involving the zero-mode
and instantaneous contributions, our LFQM result
of twist-3 DAs $\phi^P_{3;M}(x)$ and $\phi^{\perp}_{3;V}(x)$
provides the consistency with the chiral symmetry
anticipated from QCD's conformal limit~\cite{BF,Ball98}.

The purpose of this work is to extend our previous work~\cite{CJ2015} to analyze the twist-3
pseudotensor DA $\phi^\sigma_{3;\pi}(x)$ of a pion within the LFQM.
The paper is organized as follows: In Sec.~\ref{sec:II}, we compute the twist-3 pseudotensor DA $\phi^\sigma_{3;\pi}(x)$ in an exactly solvable model
based on the covariant BS model of (3+1)-dimensional fermion field theory.
We then linked the covariant BS model to the standard LFQM following
the correspondence relation~\cite{CJ2015} between the covariant BS and LFQM models
and present the form of $\phi^\sigma_{3;\pi}(x)$ as well as $\phi^P_{3;\pi}(x)$ in our
LFQM. In Sec.~\ref{sec:III}, we present our numerical results.
Summary and discussion follow in Sec.~\ref{sec:IV}.

\section{Model Description}
\label{sec:II}
%

The $\phi^P_{2;M}$ and $\phi^\sigma_{3;M}$ are defined in terms of the
following matrix elements of gauge invariant nonlocal operators
in the light-cone gauge~\cite{BF,BBL}:
\bea\label{Deq:3.1}
\phi^P_{3;M}(x) &=& \frac{2 (P\cdot\eta)}{f_M \mu_M}
\int^\infty_{-\infty}  \frac{d\tau}{2\pi} e^{-i\zeta\tau(P\cdot\eta)} \la 0|{\bar q}(\tau \eta)i\gamma_5 q(-\tau\eta)|M(P)\ra,
\\
\phi^\sigma_{3;M}(x) &=& - \frac{12}{f_M \mu_M}
\int^\infty_{-\infty}  \frac{d\tau}{2\pi} \int^x_0 dx'
e^{-i\zeta'\tau(P\cdot\eta)}
\la 0|{\bar q}(\tau\eta) i(\slash\!\!\!\!P \slash\!\!\!\eta - P\cdot\eta)\gamma_5 q(-\tau\eta)|M(P)\ra,
\eea
respectively, where $\eta=(1,0,0,-1)$  and $P$ is the four-momentum of the meson ($P^2=m^2_M$) and  $x$ corresponds to the longitudinal momentum fraction
carried by the quark and $\zeta =2x -1$.
The normalization parameter $\mu_M = m^2_M /(m_q + m_{\bar q})$
results from quark condensate. For the pion, $\mu_\pi = -2\la {\bar q}q\ra / f^2_\pi$ from the Gell-Mann-Oakes-Renner relation~\cite{GOR}.
The nonlocal matrix elements
${\cal M}_{\alpha} \equiv \la 0|{\bar q}(\tau\eta) i\Gamma_\alpha q(-\tau\eta) |M(P)\ra$
for pseudoscalar ($\Gamma_\alpha=\gamma_5$) and pseudotensor ($\Gamma_\alpha=(\slash\!\!\!\!P \slash\!\!\!\eta - P\cdot\eta)\gamma_5$) channels are
given by the following momentum integral in two-point function of the manifestly covariant BS model
\be\label{Deq:4}
{\cal M}_\alpha = N_c
\int\frac{d^4k}{(2\pi)^4} e^{-i \tau k\cdot\eta} e^{-i \tau(k-P)\cdot\eta}
\frac{{\rm Tr}\left[i\Gamma_\alpha\left(\slash \!\!\!p+m_q \right)
 \gamma_5 \left(-\slash \!\!\!k + m_{\bar q} \right) \right]} {(p^2 -m^2_q +i\varepsilon) (k^2 - m^2_{\bar q}+i\varepsilon)}
H_0,
\ee
where $N_c$ denotes the number of colors.
The quark propagators of mass $m_q$ and $m_{\bar q}$ carry the internal four-momenta $p =P -k$ and $k$, respectively.
In order to regularize the covariant loop,
we use the usual multipole ansatz~\cite{CJ_V14,CJ2015} for the $q{\bar q}$ bound-state vertex function
$H_0=H_0(p^2,k^2)$ of a meson:
$H_0(p^2,k^2) = g/(p^2 - \Lambda^2 +i\varepsilon)^2$,
where $g$ and $\Lambda$ are constant parameters.
After a little manipulation, we obtain for the pion ($m_q=m_{\bar{q}}$)
\bea\label{Deq:5.1}
\phi^P_{3;\pi}(x) &=& \frac{N_c}{f_\pi\mu_\pi} \int\frac{d^2{\bf k}_\perp}{8\pi^3}
 \frac{\chi(x,{\bf k}_\perp)}{(1-x)} M^2_0,\\
\phi^\sigma_{3;\pi}(x)
&=& -\frac{6N_c}{f_\pi\mu_\pi} \int
 \frac{d^2{\bf k}_\perp}{8\pi^3} \int^x_0 dx' \frac{(2x'-1)}{x' (1-x')}
 \frac{\chi(x',{\bf k}_\perp)}{(1-x')}  ({\bf k}^2_\perp + m^2_q),
\eea
where
$\chi(x,{\bf k}_\perp) = \frac{g}{[x (m_M^2 -M^2_0)][x (m_M^2 - M^2_{\Lambda})]^2}$
and $M^2_{0(\Lambda)} = \frac{ {\bf k}^{2}_\perp + m^2_q(\Lambda^2)}{x}
 + \frac{{\bf k}^{2}_\perp + m^2_{\bar q}}{(1-x)}$.

In the standard LFQM,
the wave function of a ground state pseudoscalar meson as a $q\bar{q}$ bound state is given by
$\Psi_{\lam{\bar\lam}}(x,{\bf k}_{\perp})
={\Phi_R(x,{\bf k}_{\perp})\cal R}_{\lam{\bar\lam}}(x,{\bf k}_{\perp})$,
where $\Phi_R$ is the radial wave function and the
spin-orbit wave function ${\cal R}_{\lam{\bar\lam}}$
with the helicity $\lam({\bar\lam})$ of a quark(antiquark)
that is obtained by the interaction-independent Melosh transformation~\cite{Melosh,Mel2}
from the ordinary spin-orbit wave function assigned by the quantum numbers $J^{PC}$.
For the radial wave function $\Phi_R$,
we use both the gaussian or harmonic oscillator (HO) wave function $\Phi_{\rm HO}$ and the power-law (PL) type
wave function $\Phi_{\rm PL}$~\cite{schlumpf} as follows
\be\label{QM2}
\Phi_{\rm HO}(x,{\bf k}_{\perp})=
\frac{4\pi^{3/4}}{\beta^{3/2}} \sqrt{\frac{\partial
k_z}{\partial x}} e^{-{\vec k}^2/2\beta^2},\;
\Phi_{\rm PL}(x,{\bf k}_{\perp})= \sqrt{\frac{128\pi}{\beta^3}}
\sqrt{\frac{\partial k_z}{\partial x}}
\frac{1}{(1 + {\vec k}^2/\beta^2)^2},
\ee
where ${\bf k}^2={\bf k}^2_\perp + k^2_z$ and $k_z=(x-1/2)M_0 +
(m^2_{\bar q}-m^2_q)/2M_0$
and $\beta$ is the variational parameter
fixed by the analysis of meson mass spectra~\cite{CJ_99}.
The normalization of $\Phi_R$ is given by
$\int\frac{dx d^2{\bf k}_\perp}{16\pi^3}
|\Phi_R(x,{\bf k}_{\perp})|^2=1$.

In our previous analyses of twist-2 and pseudoscalar twist-3 DAs of a pion~\cite{CJ2015},
we have shown that the SLF results in the standard LFQM is obtained by the
the replacement of the LF vertex function $\chi$ in BS model with our LFQM wave function
$\Phi_R$ as follows~(see Eq.~(35) in~\cite{CJ2015})
\be\label{Deq:12}
 \sqrt{2N_c} \frac{ \chi(x,{\bf k}_\perp) } {1-x}
 \to \frac{\Phi_R (x,{\bf k}_\perp) }
 {\sqrt{{\bf k}^2_\perp + m_q^2}}.
  \ee
The correspondence in Eq.~(\ref{Deq:12}) is valid again in this analysis of a
pseudotensor twist-3 DA $\phi^{\sigma}_{3;\pi}(x)$.
We now apply this correspondence to both pseudoscalar DA $\phi^{P}_{3;\pi}(x)$ in Eq.~(\ref{Deq:5.1})  and
pseudotensor DA $\phi^{\sigma}_{3;\pi}(x)$ in Eq.~(5)
to obtain them in our LFQM as follows:
\bea\label{Deq:13}
 \phi^{P}_{3;\pi}(x)
 &=& \frac{\sqrt{2N_c}}{f_\pi\mu_\pi}
 \int \frac{d^2{\bf k}_\perp}{16\pi^3}
 \frac{\Phi_R(x,{\bf k}_\perp)}{\sqrt{{\bf k}^2_\perp + m_q^2}}
 M^2_0,
 \\
 \phi^{\sigma}_{3;\pi}(x)
 &=& \frac{6\sqrt{2N_c}}{f_\pi\mu_\pi}
 \int \frac{d^2{\bf k}_\perp}{16\pi^3} \int^x_0 dx' \frac{(2x'-1)}{x' (1-x')}
 \frac{\Phi_R(x',{\bf k}_\perp)}{\sqrt{{\bf k}^2_\perp + m_q^2}}
 ({\bf k}^2_\perp  + m^2_q),
\eea
respectively.

\section{Numerical Results}
\label{sec:III}

In the numerical computations, we use $m_q=(0.22, 0.25)$ GeV and $\beta_{q\bar{q}}=(0.3659,0.3194)$ GeV for the linear (HO)~confining potential model parameters
for the Gaussian wave function, which were obtained from the calculation of meson mass spectra using the variational principle
in our LFQM~\cite{CJ_DA,CJ_99}. For the PL wave function, we use $m_q=0.25$ GeV and $\beta_{q\bar{q}}=0.335$ GeV adopted from~\cite{schlumpf}.
We first compute the second transverse moments $\la {\bf k}^2_\perp \ra^{P(\sigma)}_\pi$ for both pseudoscalar ($P$) and pseudotensor ($\sigma$) channels,
and the results obtained from the linear~[HO] parameters are given by
$\la {\bf k}^2_\perp \ra^P_\pi = (553~{\rm MeV})^2 [(480~{\rm MeV})^2]$ and
$\la {\bf k}^2_\perp \ra^\sigma_\pi = (481~{\rm MeV})^2 [(394~{\rm MeV})^2]$, respectively.

\begin{figure}
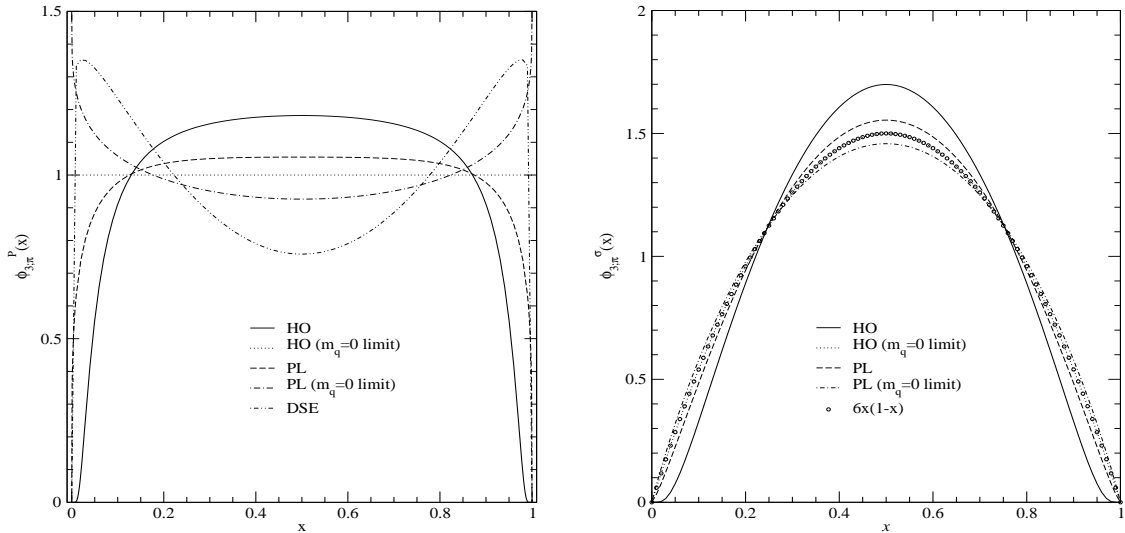

\vspace{0.5cm}
\begin{center}
\includegraphics[height=7cm, width=7cm]{Fig1a.eps}
\hspace{0.5cm}
\includegraphics[height=7cm, width=7cm]{Fig1b.eps}
\caption{\label{fig1} The twist-3 DAs $\phi^P_{3;\pi}(x)$  (left panel) and $\phi^\sigma_{3;\pi}(x)$ (right panel) of pion.}
\end{center}
\end{figure}

Fig.~\ref{fig1} shows the two-particle twist-3 pion $\phi^P_{3;\pi}(x)$  (left panel) and $\phi^\sigma_{3;\pi}(x)$ (right panel)
obtained from  the nonzero constituent quark masses using Gaussian wave functions with HO (solid lines) model parameters
and PL wave functions (dashed lines). We also plot our results in the chiral symmetry $(m_{u(d)}\to 0)$ limit for both Gaussian (dotted lines)
and PL (dot-dashed lines) wave functions and compare them with the chiral-limit prediction of DSE approach
employing the dynamical chiral symmetry breaking-improved kernels~\cite{DSE2015} (double-dot-dashed line) as well as
the asymptotic result $6x(1-x)$ (circled data) for the case $\phi^\sigma_{3;\pi}(x)$.
Our results  for both $\phi^P_{3;\pi}(x)$  and $\phi^\sigma_{3;\pi}(x)$  are normalized without the momentum cutoff (i.e. $|{\bf k}_\perp|\to\infty$).

For $\phi^P_{3;\pi}(x)$ case, our results with nonzero constituent quark masses show rather convex shapes for both Gaussian and PL
wave functions but they show quite different end point behaviors, i.e. the end points are more enhanced for the PL wave function than
the Gaussian wave function. The difference between the two wave functions are more drastic in the chiral symmetry limit, where the result of Gaussian
wave function  reproduces the  result $\phi^P_{3;\pi}(x)\to 1$  anticipated from the QCD's conformal limit~\cite{BF}
but the result of PL wave function shows the concave shape similar to the result of DSE approach~\cite{DSE2015}, in which
the following asymptotic form $\phi^P_{3;\pi}(x)\to 1 + (1/2) C^{(1/2)}_2(2x-1)$ was obtained.
While the authors in~\cite{DSE2015} explained that the difference, i.e. $(1/2)C^{(1/2)}_2(2x-1)$ term in chiral symmetry limit may come from the
mixing effect between the two- and three-particle twist-3 amplitudes, we observe that this difference may come  from the cutoff scale of the transverse
momentum scale associated with the different choice of LF wave functions.
 Especially, 
as on can see from Fig. 1, while the end-point suppressed form of $\phi^P_{3;\pi}(x)$ obtained from HO wave function shows
constant shape (dotted line) in the chiral symmetry limit, the end-point enhanced form of $\phi^P_{3;\pi}(x)$ obtained from PL wave function
shows concave shape (dot-dashed line) in the chiral symmetry limit.
The cutoff dependent behaviors of $\phi^P_{3;\pi}(x)$ obtained from both Gaussian and PL wave functions are also presented in~\cite{FB2015}, where
the concave shape for the Gaussian wave function can also be seen with the cutoff scale $\mu=1$ GeV or less being taken but the
cutoff dependence was shown to be more sensitive for the PL wave function than the Gaussian one.

For $\phi^\sigma_{3;\pi}(x)$ case,  our results with nonzero constituent quark masses for both Gaussian (solid line) and PL (dashed line)
show again different end point behaviors, i.e. the end points are more enhanced for the PL wave function than
the Gaussian wave function.
However, in the chiral symmetry limit,  Gaussian (dotted line) and PL (dot-dashed line) wave functions show very
similar shapes each other. Furthermore, the result from Gaussian wave function reproduces exactly the asymptotic form $6x(1-x)$.
The same chiral-limit behavior was also obtained from the DSE approach~\cite{DSE2015}.
As one can see from Fig.~\ref{fig3}, the twist-3 pseudoscalar $\phi^P_{3;\pi}(x)$
is more sensitive to the shape of
the model wave functions (Gaussian vs. PL) than the twist-3 pseudotensor $\phi^\sigma_{3;\pi}(x)$.
It is quite interesting to note in the chiral symmetry limit that while $\phi^P_{3;\pi}(x)$ is sensitive to the shapes of model wave functions,
$\phi^\sigma_{3;\pi}(x)$ is insensitive to them.

The twist-3 pseudoscalar DA $\phi^P_{3;M}(x)$ and pseudotensor DA $\phi^\sigma_{3;M}(x)$ are usually expanded in terms of the
Gegenbauer polynomials $C^{1/2}_n$ and $C^{3/2}_n$, respectively, as follows~\cite{NK06}:
$\phi^P_{3;M} =  \sum^{\infty}_{n=0} a^{P}_{n,M}C^{1/2}_n(2x-1)$,
and
 $\phi^\sigma_{3;M} = 6 x(1-x) \sum^{\infty}_{n=0} a^{\sigma}_{n,M}C^{3/2}_n(2x-1)$.
The coefficients
$a^{P(\sigma)}_{n,M}$ are called the Gegenbauer moments,  which describe how much the DAs deviate from the asymptotic one.
In addition to the Gegenbauer moments, one can also define the expectation value of the longitudinal
momentum, so-called $\xi$-moments, as follows:
$\la\xi^n\ra^{P(\sigma)}_M = \int^1_0 dx \xi^n \phi^{P(\sigma)}_{3;M}(x)$.
The Gegenbauer- and $\xi$- moments of the pseudoscalar twist-3 DAs
$\phi^P_{3;\pi}(x)$ and twist-2 DA $\phi^A_{2;\pi}(x)$ can be found
in our previous works~\cite{CJ_DA,CJ2015}.

\begin{table}[t]
\caption{The Gegenbauer moments and $\xi$ moments of  twist-3 pion DAs obtained from the
linear and HO potential models compared other model estimates. }
\label{t1}
\renewcommand{\tabcolsep}{1pc} 
\centering
\begin{tabular}{@{}lcccccc}  \hline\noalign{\smallskip}
Models &  $a^{\sigma}_{2,\pi}$ & $a^{\sigma}_{4,\pi}$ & $a^{\sigma}_{6,\pi}$
& $\la\xi^2\ra^{\sigma}_\pi$ & $\la\xi^4\ra^{\sigma}_\pi$ & $\la\xi^6\ra^{\sigma}_\pi$ \\[3pt]
\tableheadseprule\noalign{\smallskip}
HO & -0.1155 & -0.0268 & -0.0046 & 0.1604 & 0.0565 & 0.0263 \\
Linear & -0.0803 &  -0.0256 &  -0.0082 & 0.1725 &  0.0647 & 0.0318 \\
PL       & -0.0375 & -0.0092 & -0.0031 & 0.1871 & 0.0762 & 0.0406 \\
SR~\cite{BBL} & 0.0979&  -0.0016 &  -0.0011 & 0.2325 &  0.1075 & 0.0624 \\
DSE~\cite{DSE2015} & $\cdots$ &  $\cdots$ &  $\cdots$ & 0.20 & 0.085 & 0.047 \\
$\chi$QM~\cite{NK06} & -0.0984 &  -0.0192 &  -0.0037 & 0.1663 &  0.0612 & -0.0015 \\
$6x(1-x)$ & $\cdots$ &  $\cdots$ &  $\cdots$ & 0.20 & 0.086 & 0.048 \\
\noalign{\smallskip}\hline
\end{tabular}
\end{table}

In Table~\ref{t1}, we list the calculated Gegenbauer- and $\xi$- moments of  pseudotensor twist-3 pion DA
$\phi^\sigma_{3;\pi}(x)$
obtained from the Gaussian wave functions
with linear and HO potential models and PL wave function.
We also compare our results  with other model predictions, e.g.
QCD sum rules (SR)~\cite{BBL}, DSE approach~\cite{DSE2015} and the chiral quark model ($\chi$QM)~\cite{NK06}.
As expected from the isospin symmetry, all odd Gegenbauer and $\xi$ moments are all zero.
It is interesting to note that the sign of
$a^{\sigma}_{2,\pi}$ is negative from our LFQM and $\chi$QM predictions but is positive for QCDSR prediction.
Larger positive value of $a^{\sigma}_{2,\pi}$ leads to more flat shape of DA but the larger negative value leads to more narrower
shape of DA.
Knowing our LFQM results from the HO model are exact to the asymptotic result in the chiral-symmetry limit,
i.e. $[\la\xi^2\ra^{\sigma}_\pi]_{\rm HO}=[\la\xi^2\ra^{\sigma}_\pi]_{\rm asy}$ in $m_q\to 0$ limit, one can see
that the $\xi$-moments are reduced when the chiral symmetry is broken. We also should note
for the same reason
that our LFQM results are in good agreement with DSE results in the chiral-symmetry limit of $\phi^\sigma_{3;\pi}(x)$.

\section{Summary}
\label{sec:IV}

We analyzed the two twist-3 DAs of pion, i.e. pseudoscalar $\phi^P_{3;\pi}(x)$
and pseudotensor $\phi^\sigma_{3;\pi}(x)$, within the LFQM. We also
investigated the discrepancy of the asymptotic forms of $\phi^P_{3;\pi}(x)$ between
DSE approach~\cite{DSE2015} and QCD's conformal limit expression~\cite{BF} from the perspective of
dependence of DA on the form of LF wave functions, e.g. Gaussian vs. PL wave functions.
In order to compute the twist-3 pseudotensor DA $\phi^\sigma_{3;\pi}(x)$,
we utilized the same manifestly covariant BS model used in~\cite{CJ_V14,CJ2015,FB2015} and then substituted the LF vertex function in the covariant BS model
with the more phenomenologically accessible Gaussian and PL wave functions.
Linking the covariant BS model to the standard LFQM, we used the same correspondence
relation Eq.~(\ref{Deq:12}) between the two models as the one found in~\cite{CJ2015,CJ_V14}.
The remarkable finding in linking the covariant BS model to the standard LFQM
is that the treacherous points such as the zero-mode contributions and the instantaneous
ones existed in the covariant BS model become absent in the LFQM with the Gaussian
or PL wave function.

Our LFQM descriptions for both twist-3 $\phi^P_{3;\pi}$  and  $\phi^\sigma_{3;\pi}$
satisfy the fundamental constraint(i.e. symmetric form with respect to $x$) anticipated from the isospin symmetry. For the $\phi^P_{3;\pi}(x)$ case, our results with nonzero constituent quark masses show rather convex shapes for both Gaussian and PL
wave functions but they show quite different end point behaviors, i.e. the end points are more enhanced for the PL wave function than
the Gaussian wave function. The difference between the two wave functions are more drastic in the chiral symmetry limit, where the result of Gaussian
wave function  reproduces the  result $\phi^P_{3;\pi}(x)\to 1$  anticipated from the QCD's conformal limit~\cite{BF}
but the result of PL wave function shows the concave shape similar to the result of DSE approach~\cite{DSE2015}.
For the $\phi^\sigma_{3;\pi}(x)$ case,  our results from both Gaussian and PL wave functions in the chiral symmetry limit, show very similar shapes each other.
Especially, the result from Gaussian wave function reproduces exactly the asymptotic form $6x(1-x)$ anticipated from QCD's conformal limit.
The same chiral-limit behavior was also obtained from the DSE approach~\cite{DSE2015}.
The remarkable thing is that our predictions for the two twist-3 DAs of $\pi$
and chirality-even twist-2  DA $\phi^{||}_{2;\rho}(x)$ and twist-3 DA $\phi^{\perp}_{3;\rho}(x)$ of $\rho$~\cite{CJ_V14} obtained from the Gaussian
wave function in the chiral limit exactly reproduce the forms anticipated from QCD's conformal limit.
We summarize in Table~\ref{t2} the asymptotic forms for the two twist-3 DAs of $\pi$
and chirality-even twist-2  and twist-3 DAs of rho meson compared with DSE approach~\cite{DSE2015} and QCD's conformal limit expression~\cite{BF,Ball98}.

\begin{table}[t]
\caption{Asymptotic forms of various DAs of $\pi$ and $\rho$.}
\centering
\label{t2}
\begin{tabular}{lllll} \hline\noalign{\smallskip}
Model in $m_q\to 0$  & $\phi^P_{3;\pi}$ & $\phi^\sigma_{3;\pi}$ & $\phi^{||}_{2;\rho}(x)$ & $\phi^{\perp}_{3;\rho}(x)$\\[3pt]
\tableheadseprule\noalign{\smallskip}
LFQM (HO model)~\cite{CJ_V14,CJ2015} &  1 &  $6x(1-x)$ & $6x(1-x)$ & $\frac{3}{4}[1+ (2x-1)^2]$  \\
DSE approach~\cite{DSE2015} &  $1+ \frac{1}{2} C^{1/2}{2}(2x-1)$ & $6x(1-x)$ & $-$ & $-$  \\
QCD's conformal limit ~\cite{BF,Ball98}& 1&  $6x(1-x)$ & $6x(1-x)$ & $\frac{3}{4}[1+ (2x-1)^2]$  \\
\noalign{\smallskip}\hline
\end{tabular}
\end{table}

The idea of our LFQM is to provide the nonperturbative wave functions at the momentum scale
consistent with the use of constituent quark mass. The DAs determined from this nonperturbative wave functions can be fed into the QCD evolution equation to provide the shorter distance information of the corresponding hadrons. 
While our results for both $\phi^P_{3;\pi}$ and $\phi^\sigma_{3;\pi}$ in Fig. 1 were obtained without 
the momentum cutoffs (i.e. $|{\bf k}_\perp|\to\infty$), the scales of both DAs obtained from the Gaussian wave functions 
and that of $\phi^\sigma_{3;\pi}$ obtained from PL wave function are estimated to be $|{\bf k}_\perp|\simeq 1$ GeV.
However, the scale of $\phi^P_{3;\pi}$ obtained from PL wave function are estimated to  $|{\bf k}_\perp|\simeq 2$ GeV due to the high momentum tail compared to
other wave functions.  
The DAs obtained without the cutoff should not be regarded as the fully evolved DAs but still be nonperturbative as they
just mean that the cutoff dependence becomes marginal beyond a certain nonperturbative cutoff scale.

\end{document}